\documentclass[aps,prb,twocolumn]{revtex4}

\usepackage{graphics}

\begin{document}

\preprint{p01jan04}

\title{Hiawatha's Valence Bonding}

\author{R. B. Laughlin}

\affiliation{Department of Physics, Stanford University, Stanford,
             CA 94305}

\homepage[ \copyright 2004 R. B. Laughlin.  This work is distributed under
the terms of a Creative Commons license.  The author grants permission to
copy, distribute, display, and perform the work in unaltered form, with
attribution to the author, for noncommercial purposes only. All 
other rights, including commercial rights, are reserved to the author.  ] 
{http://large.stanford.edu}

\date{January 1, 2004}

\begin{abstract}

There is increasing circumstantial evidence that the cuprate
superconductors, and correlated-electron materials generally, defy simple
materials categorization because of their proximity to one or more
continuous zero-temperature phase transitions.  This implies that the
fifteen-year confusion about the cuprates is not fundamental at all but
simply overinterpreted quantum criticality---an effect that seems
mysterious by virtue of its hypersensitivity to perturbations, {\it i.e.}
to sample imperfections in experiment and small modifications of
approximation schemes in theoretical modeling, but is really just an
unremarkable phase transition of some kind masquerading as something
important, a sheep in wolf's clothing.  This conclusion is extremely
difficult for most physicists even to think about because it requires
admitting that an identifiable physical phenomenon might cause the
scientific method to fail in some cases.  For this reason I have decided
to explain the problem in a way that is nonthreatening, easy to read, and
fun---as a satire modeled after a similar piece of Lewis Carroll's I once
read.  My story is humorous fiction. Any similarity of the characters to
living persons is accidental. My apologies to Henry W. Longfellow.
[Published as Annals of Improbable Research {\bf 10}, No. 6 (May/June 
2004), p. 8.]

\end{abstract}

\maketitle

\section{Introduction}

\begin{verse}
Since all men have imperfections\\
Hanging bones inside their closets\\
That they trust no one will notice\\
Absent tips on where to find them,\\
It will shock no one to learn that\\
Even mighty Hiawatha\\
Famous Chief of myth and legend\\
Did some things he was not proud of\\
While a brother in a frat house\\
With a surly reputation\\
At an unknown little college\\
That his father helped to finance\\
So that he would get admitted\\
By the shores of Gitche-Gumee.\\
\end{verse}

\begin{verse}
Far from loving fields and flowers\\
And the odor of the forest\\
As one reads in all the textbooks\\
Hiawatha hated woodlands\\
And the animals one finds there,\\
Whom he felt were always pooping,\\
And the plants the critters fed on\\
Down in dank and swampy bottoms,\\
Nearly perfect grounds for breeding\\
Mighty hordes of great mosquitoes\\
Who were always lean and hungry\\
And equipped with maps and radar\\
Could detect where you were hiding\\
To inflict their bites and torments,\\
With their sneaky friends the black flies,\\
And their angry friends the green flies,\\
And the rocks ensnared by tree roots\\
That existed just to trip you\\
And would look improved as concrete\\
In foundation for a condo.\\
\end{verse}

\begin{verse}
Thus the kindly, thoughtful image\\
Of a noble man of Nature\\
Was a total fabrication\\
Of a team of gifted spin docs\\
Hired later for this purpose.\\
He was really just a tech nerd\\
Who cared only for equations\\
And explaining all behavior\\
From the basic laws of physics\\
Armed with only mathematics.\\
\end{verse}

\begin{figure}[t!]
\begin{center}
\includegraphics{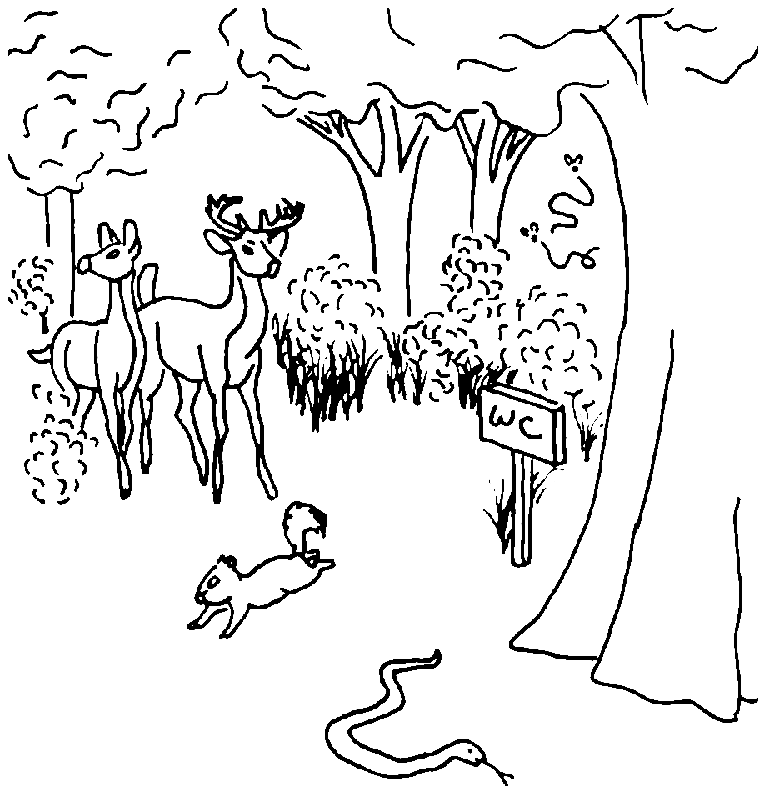}
\end{center}
\begin{center}
{\it Hiawatha hated woodlands.}
\end{center}
\end{figure}

\begin{verse}
Thus, instead of lakes and forests,\\
Hiawatha worshipped Newton,\\
Whose account of Kepler's orbits\\
Built on rules that Galileo\\
Had inferred from observation\\
Plus the innocent assumption\\
Of a law of gravitation\\
Was a cosmic inspiration;\\
And the brilliant Sadie Carnot,\\
Whose insightful laws of heat flow\\ 
Were deduced from working engines\\
Absent microscopic theories;\\
And the tragic Ludwig Boltzmann\\
Who ascribed these laws to counting\\
But fell victim to depression\\
When he found no one believed him\\
And so killed himself by jumping\\ 
From an Adriatic tower.\\
Hiawatha saw that Maxwell's\\
Guessing missing laws of motion\\
Needed for predicting light waves,\\
Was the most transcendent genius,\\
As was Albert Einstein's insight\\
That the speed of light being constant\\
Must mean time was not consistent\\
And that mass could be converted\\
Into heat and vice versa.\\
Just as clear was that the Planck law\\ 
Must imply DeBroglie's wavelength\\
Was in force in any matter\\
So that sharp atomic spectra\\
And distinct atomic sizes\\
And the laws of bond formation\\
Came from quantum interference.\\
\end{verse}

\section{Hiawatha's Mistake}

\begin{verse}
Thus it was that Hiawatha\\
Came to be infatuated\\
With the laws of quantum matter,\\
Which means liquid noble gases,\\
Neutrons in a burnt-out star core,\\
Or just rocks so cryogenic\\
They cannot get any colder,\\
Even with improved equipment,\\
Like the state of too much sliding\\
On the ice of Gitche-Gumee\\
After dark in dead of winter\\
In an inexpensive loincloth.\\
Pain and danger notwithstanding\\
Quantum matter's simple structure\\
Makes the eager physics tyro\\
Quite unable to resist it.\\
Hiawatha learned how atoms\\
Self-assemble into crystals,\\
How electrons move right through them,\\
Waving past the rigid ions\\
Thereby making them metallic\\
In the absence of a bandgap\\
Which arises from diffraction\\
And prevents the charge from moving\\
Thereby causing insulation,\\
But by means of wires and doping\\
With atomic imperfections\\
When the bandgap is a small one\\
Can be used to make transistors.\\
In addition to the basics\\
He learned how electric forces\\
Like those seen in clinging woolens\\
Cause some things to be magnetic\\
Up until the lowly phonon,\\
Quantum particle of sound wave,\\
Storing heat the way that light does\\
Mediates a strong attraction\\
That can pair up two electrons\\
Causing them to move together\\
Overcoming all resistance\\
And producing other magic\\
Such as quantum oscillations.\\
\end{verse}

\begin{figure}[t!]
\begin{center}
\includegraphics{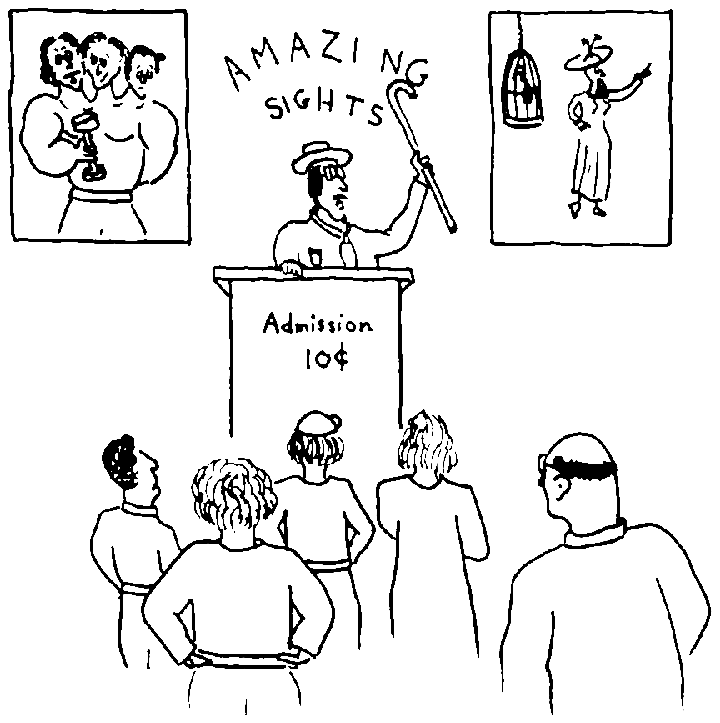}
\end{center}
\begin{center}
{\it They were little more than con men. }
\end{center}
\end{figure}

\begin{verse}
At this quite untimely moment\\
Of his fragile student history\\
When his mind was most suggestive\\
Our poor hapless Hiawatha\\
Had the terrible misfortune\\
To fall in with wicked people\\
Who were little more than con men\\
And advanced in their profession\\
Making theories of such matter\\
That were not at all deductive\\
But instead used mathematics\\
As a way to sow confusion\\
So that no one would discover\\
That their stuff was pure opinion\\
Spiced with politics and chutzpah\\
So it looked somewhat like science\\
Even though it really wasn't.\\
\end{verse}

\begin{verse}
How they did this was ingenious\\
For it's not a simple matter\\
To produce concrete equations\\
That are absolutely hokum\\
And escape without detection\\
When they represent relations\\
Of some quantities one measures\\
Written down as abstract symbols\\
That could easily be tested.\\
What they did was deftly prey on\\
Prejudicial ways of thinking\\
That their colleagues thought were reasoned\\
But were simply misconceptions,\\
Generated during training\\
They had all received as students,\\
That the properties one wanted\\
Were completely universal\\
So details did not matter.\\
But the data did not say this\\
And, moreover, had they done so\\
There would have been no good reason\\
To think any more about it.\\
So, while everyone was watching,\\
They swapped in some new equations\\
That they said would solve the problem\\
On account of being much simpler\\
But in fact described a system\\
Very different from the first one\\
And, moreover, was unstable,\\
Balanced at competing phases,\\
So that nobody could solve it\\
Thus betraying the deception.\\
\end{verse}

\begin{verse}
Adding to the dazzling brilliance\\
Of this coldly thought-out swindle\\
They declared it {\it fundamental}\\
So that all the strange creations\\
Made by people trying to solve it\\
And quite clearly not succeeding\\
Proved it was a fount of deepness\\
One should struggle to unravel\\
Even if it took a lifetime.\\
As a nifty added bonus\\
Any hint you dropped in public\\
That it might have no solution\\
Simply meant you weren't a genius,\\
Told the world that you were stupid,\\
That you were a hopeless failure\\
Who should not command a pencil.\\
No one wanted to admit this\\
So they'd cover up their failure\\
And pretend that they had solved it\\
Even though they clearly hadn't.\\
This succeeded, for the most part,\\
But in one respect it didn't,\\
For their desperate need to publish\\
And thereby maintain their funding\\
Caused a massive flood of papers,\\
Each quite different from the others,\\
To descend upon the journals\\
And to overwhelm and clog them.\\
This would have been very funny\\
Had it not been so pathetic.\\
\end{verse}

\begin{figure}[t!]
\begin{center}
\includegraphics{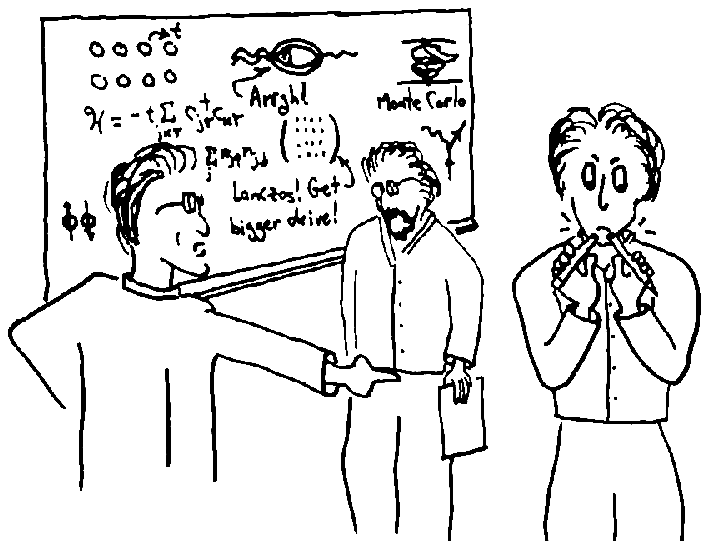}
\end{center}
\begin{center}
{\it It just meant you weren't a genius.}
\end{center}
\end{figure}

\begin{verse}
Hiawatha bought the story\\
Took the bait, hook, line, and sinker\\
And, like many other students\\
Who'd been victimized before him,\\
Got convinced that his strong math skills,\\
Far exceeding those of others,\\
Would reveal nature's mysteries\\
When he solved the Hubbard model\\
And its child the t-J model\\
And the lattice Kondo model\\
And the quantum spin glass model,\\
All of which possessed the feature\\
That no human being could solve them.\\
\end{verse}

\section{Hiawatha Meets the Cuprates}

\begin{verse}
Nature has a sense of humor,\\
As one learns by working with it,\\
But it is an opportunist,\\
So that life's most bitter lessons\\
Often wind up learned the hard way\\
When it moves to take advantage\\
Of a single bad decision\\
And compound it with some mischief\\
Custom made for the occasion.\\
\end{verse}

\begin{verse}
Just when he'd resolved to strike out\\
On his suicidal mission\\
There occurred a bold announcement\\
In a well-known German journal\\
That a tiny lab near Z\"{u}rich\\
Had discovered a material\\
With the structure of perovskite\\
Made of oxygen and copper\\
And some other stuff like strontium\\
That when cooled to thirty kelvin\\
Lost all traces of resistance.\\
This event was simply shocking\\
For existing quantum theory\\
Said it had to get much colder\\
For this special thing to happen,\\
As did all the careful surveys\\
Of the properties of metals,\\
Which were very comprehensive\\
And agreed well with the theory.\\
Since the chemists were ambitious\\
To somehow transcend this limit,\\
Which they thought too academic,\\
And someday kill all resistance\\
Using no refrigeration,\\
There ensued a feeding frenzy\\
Worthy of a horror movie,\\
Like what happens when a trawler\\
Dumps its hold of tuna entrails\\
Off a reef in north Australia.\\
\end{verse}

\begin{verse}
One example of this madness\\
Was the {\it Physics Woodstock} conference\\
That took place in mid-Manhattan\\
Shortly after the announcement\\
Where attendees got together,\\
Comandeered a giant ballroom,\\
And gave talks not on the program\\
In a special all-night session\\
Dedicated to the cuprates\\
Which was packed to overflowing.\\
There was talk of maglev transport,\\
New kinds of computer circuit,\\
Mighty, compact little motors\\
And efficient power cables,\\
All of which would soon be coming\\
Thanks to this momentous breakthrough.\\
But it turns out we don't have them\\
For they weren't a big improvement\\
Over things we had already\\
And were hopelessly expensive.\\
\end{verse}

\begin{verse}
Then there were the frantic searches\\
To find compounds that were better,\\
Which one knew could be accomplished\\
If one spent enough time looking,\\
Since this stuff had lots of phases\\
Subtly different from each other,\\
And there had to be a best one.\\
There was very rapid progress\\
Culminating in a patent\\
For a more complex material\\
In the same broad class of structure\\
Which performed at ninety kelvin,\\
So much higher than the theory\\
Would allow to ever happen\\
Even with extreme assumptions\\
That one knew it was in trouble.\\
\end{verse}

\begin{verse}
Almost overnight one found that\\
Every spectrum known to science\\
Had been taken on a cuprate.\\
Their alleged profound importance\\
Was, of course, a major factor,\\
But what mattered most was tactics.\\
Without need to tell one's funders,\\
Since it could be done so quickly,\\
One could telephone a chemist,\\
Cut a deal to get some samples,\\
Put them in one's apparatus---\\
Presto! Out would come a paper\\
That would instantly get published\\
Even if it was a stinker.\\
This produced a pile of data,\\
Growing without bound, like cancer,\\
That completely overwhelmed you\\
By being mostly unimportant,\\
Like the growing list of options\\
Coming from your cable service.\\
\end{verse}

\begin{figure}[t]
\begin{center}
\includegraphics{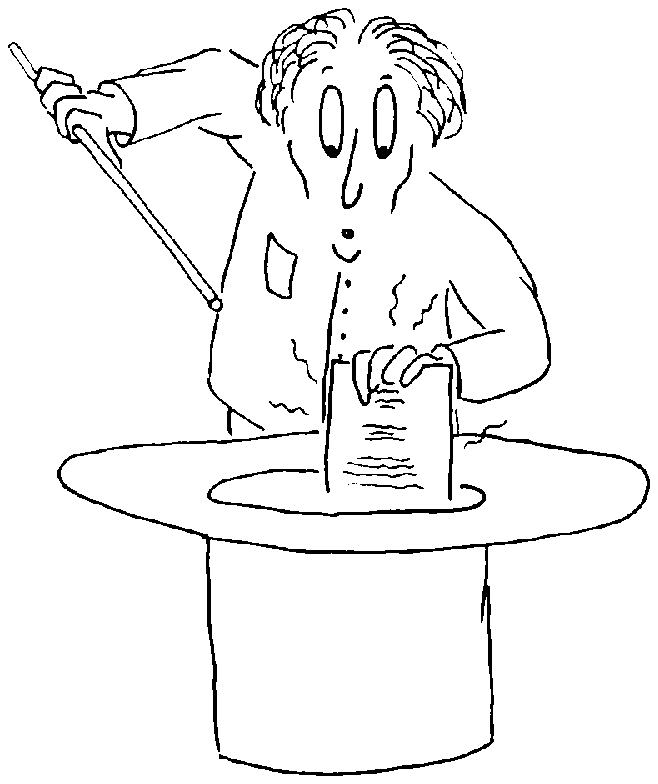}
\end{center}
\begin{center}
{\it Presto! Out would come a stinker.}
\end{center}
\end{figure}

\begin{verse}
Often spectra weren't consistent,\\
But, instead of getting angry\\
As one would have in the old days,\\
One would handle it maturely\\
And just chalk it up to errors\\
That occur when one is hasty\\
Or has had bad luck with samples.\\
But this tolerance, it turns out,\\
Was a bargain with the devil\\
For it later was discovered\\
That enormous variation\\
Was endemic to the cuprates,\\
And that things not reproducing\\
Due to complex phase inclusions,\\
Foreign atoms in the sample,\\
Careless oxygen annealing,\\
Surface preparation methods,\\
And a thousand other factors\\
Was essential to their nature.\\
\end{verse}

\begin{verse}
Sadly, by the time this surfaced\\
Shameful habits of denying\\
That the differences existed\\
Had become enshrined in writing,\\
And so wedded to the culture,\\
That they could not be corrected.\\
It was now accepted practice\\
In a public presentation\\
Of experimental findings\\
Not to mention other data\\
Even if your own group took them.\\
Grounds for this were rarely stated,\\
Other than the innuendo\\
That one's sorry competition\\
Were a hopeless bunch of bozos\\
Who did not know how to measure\\
And therefore could not be trusted.\\
It was likewise viewed as kosher\\
To make up a little theory\\
Or adopt somebody else's\\
That gave all your findings meaning---\\
Although not those of your colleagues,\\
Which were, sadly, so imperfect\\
They were simply inconsistent.\\
But one never heard recanting,\\
Since it would have meant admission\\
That one's judgement had been faulty.\\
\end{verse}

\begin{verse}
Thus the cuprates' weird caprices\\
Long escaping understanding\\
Transformed into pseudotheories\\
That, like gods on Mount Olympus,\\
Were political creations\\
That could not be killed with reason\\
And, empowered as immortals,\\
Took control of their creators,\\
Warred among themselves for power,\\
Schemed to have a lot of children,\\
And, in general, made a circus\\
Of the scientific method.\\
\end{verse}

\begin{verse}
Hiawatha, being a student,\\
And, quite frankly, rather callow\\
Did not have the slightest inkling\\
That such nonsense ever happened.\\
He believed the claims of science\\
To be rather more objective\\
Than competing kinds of knowledge\\
On account of its precision\\
And the fact that you could test it.\\
Rather than the yawning snake pit\\
Seething with disinformation\\
That was really there before him,\\
Certain death for young beginners,\\
He saw just a chance for glory\\
Something of immense importance,\\
Judging from the acrimony\\
Coursing through the talks and papers,\\
And a vast supply of data\\
On which one could build a theory\\
And thereby become a hero,\\
Much the way the dumber brother\\
Of the famous brave Odysseus\\
That no one has ever heard of,\\
Sure he could outfox the sirens,\\
Ordered that the men unbind him\\
And, of course, succumbing quickly\\
Dove right in and bashed his brains out.\\
\end{verse}

\section{Hiawatha Escapes Reality}

\begin{figure}[t!]
\begin{center}
\includegraphics{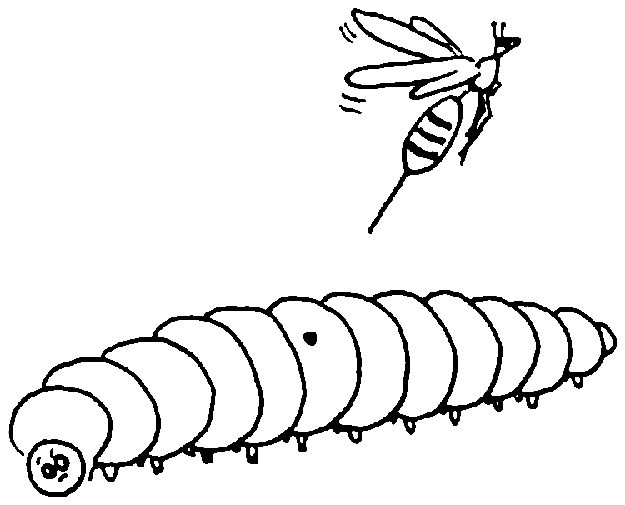}
\end{center}
\begin{center}
{\it Hiawatha's misconceptions were not shared by everybody. }
\end{center}
\end{figure}

\begin{verse}
Hiawatha's misconceptions\\
Of the nature of the problem\\
He was setting out to conquer\\
Were not shared by everybody.\\
Just as buzzards, with keen noses,\\
Circling high above their breakfast\\
Wait until it cannot hurt them,\\
To swoop down and get to business,\\
And ichneuman wasps impregnate\\
Larval caterpillar victims\\
With some eggs that grow to eat them,\\
Thus not let them reach adulthood\\
When they might be hard to handle,\\
Hiawatha's crafty mentors\\
Sensed that science had stopped working\\
In the sub-field of the cuprates,\\
As it had before in others\\
Where their scams had been successful.\\
Smelling death was close upon it,\\
They resolved the time was ready.\\
\end{verse}

\begin{verse}
What ensued was simply awesome,\\
Destined to go down in legend.\\
They proposed a cuprate theory\\
So magnificent in concept,\\
So much bolder than the others\\
That it blasted them to pieces\\
Like some big atomic warhead,\\
So outshined them in its glory\\
Like a nova in the heavens\\
That it blinded any person\\
Who would dare to gaze upon it.\\
Cuprates did these things, it stated,\\
Just because a quirk of nature\\
Made them like the {\it Hubbard model},\\
Which, as had been long established,\\
Did some things quite fundamental,\\
Not yet known to modern science,\\
Which explained the crazy data,\\
So to understand the cuprates\\
One would have to solve this model.\\
How colossal! How stupendous!\\
It was absolutely foolproof!\\
No one could disprove this theory\\
With existing mathematics\\
Or experimental data\\
For exactly the same reasons\\
Nor could they admit they couldn't,\\
So they'd spend their whole lives trying,\\
Blame themselves for being so stupid,\\
And pay homage in each paper\\
With the requisite citation!\\
\end{verse}

\begin{verse}
They left clues in great abundance\\
That they'd made a vast deception\\
Far surpassing P. T. Barnum's\\
Most creative whims and musings\\
Trusting that no one would catch them\\
On account of being so guileless,\\
Which they knew was part of science,\\
Rather like the clever killer,\\
Sure he can outsmart Columbo,\\
Leaving marks upon the crime scene\\
Then in later verbal sparring\\
Hints at them in brazen taunting.\\
One was that its short description,\\
Resonating bonds of valence,\\
Was the name that Linus Pauling\\
Used for common bonds of benzene,\\
Something so profoundly different\\
From the physics of the cuprates\\
That its use on this occasion\\
Seemed to show a lousy word sense.\\
But, in fact, it was inspired,\\
For the permanent confusion\\
Left by its uncertain meaning\\
Like the data it reflected,\\
Was defense against attackers,\\
Made it very hard to target,\\
Left its enemies bewildered.\\
And the thoughtful usurpation\\
Of a well-established brand name\\
Had the lovely added feature\\
Of dispatching pesky Pauling,\\
Who had always been a nuisance,\\
Down to Davy Jones's locker\\
In the minds of younger people.\\

\begin{figure}[t]
\begin{center}
\includegraphics{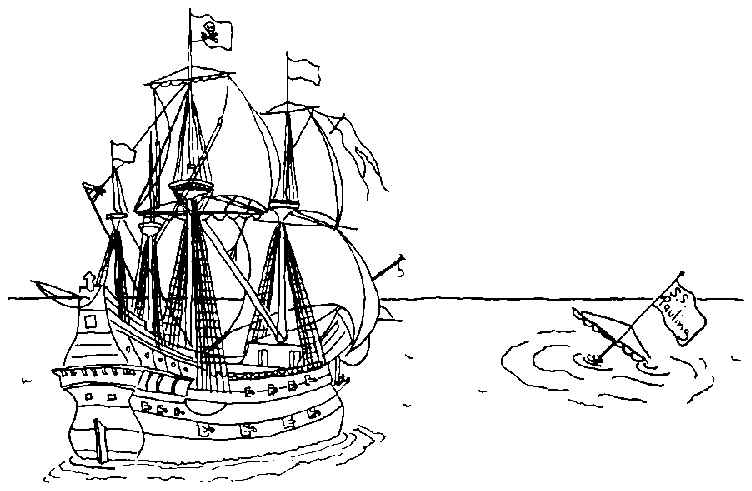}
\end{center}
\begin{center}
{\it Getting rid of pesky Pauling.}
\end{center}
\end{figure}

There was also the assertion\\
Running rampant through the theory\\
That the essence of the cuprates\\
Was coulombic insulation,\\
Which, on close inspection, turned out\\
No one could define precisely,\\
With a few concrete equations,\\
But was nonetheless a concept\\
People thought they comprehended,\\
Like the fancy secret contents\\
Of competing brands of toothpaste\\
That, of course, are total fictions\\
Made up during lunch by ad guys.\\

But the best clue by some margin\\
Was the {\it deus ex machina}\\
Known as Gutzwiller Projection,\\
Which began life as a method\\
For controlling the equations\\
But was morphed on this occasion\\
To a monsterous distortion\\
Of the basic mathematics\\
On the grounds it was insightful.\\
But, in fact, it came from nowhere,\\
And was just a simple dictat\\
That an off-the-shelf conductor\\
Could not be a quantum magnet\\
While one forced it to become one\\
Thus creating awful conflict\\
When, in fact, there simply was none.\\
\end{verse}

\begin{verse}
Hiawatha, being clever,\\
Quickly saw that he could do this,\\
Saw that such manipulations\\
Were, in fact, extremely easy,\\
That a high school kid could do them,\\
Once he got the key idea\\
That one should evade the problem\\
Of deducing the behavior\\
From the actual equations\\
By declaring that some answer\\
Was correct because one said so\\
And proceeding to defend it\\
With a lot of complex symbols\\
Simply cooked up to confuse things.\\

Thus emboldened to abandon\\
His perverse outdated fear of\\
Uncontrolled approximations\\
Hiawatha bit the bullet\\
And jumped into cuprate theory\\
With the fury of a madman,\\
Doing reckless calculations\\
Based on nothing but some gas fumes\\
That produced some fine predictions,\\
As one was inclined to call them,\\
Matching some existing data\\
But, of course, not matching others,\\
Since they were not all consistent.\\
He would then just pick and choose them\\
As one would an orange or lemon\\
In the local supermarket\\
And declare the rest defective.\\
Then he wrote up his conclusions\\
In a little physics paper\\
Loaded up with fearsome symbols\\
Proving that he had credentials\\
To make all these speculations,\\
Sent it in for publication\\
And then found an awful problem\\
He had not anticipated.\\
For the paper to be published\\
It must get past refereeing\\
Which, in theory, was for stopping\\
False results from being reported\\
But, in practice, was to censor\\
Anyone whose work you hated,\\
Somewhat of a sticky wicket\\
For someone who's main objective\\
Was to publish speculation.\\
Hiawatha soon discovered\\
Though the process of rejection\\
That his papers could not make it\\
If they championed new ideas\\
Or in any way conflicted\\
With the viewpoints of the experts\\
Which, of course, were simply made up.\\
\end{verse}

\begin{verse}
Thus the mighty Hiawatha\\
Found his plans to be a scholar\\
Had an unexpected down side\\
That would later prove quite fatal\\
In that he was forced to pander\\
In his writing for the public\\
To a set of flakey concepts\\
That he'd found extremely useful\\
But had not had time to question,\\
In exchange for recognition\\
Needed for career advancement.\\
For a while it did not matter\\
But the problem slowly festered\\
And one day poor Hiawatha,\\
Waking to a huge disaster,\\
Found himself up to his eyeballs\\
In a soup of black corruption.\\
\end{verse}

\begin{figure}[t]
\begin{center}
\includegraphics{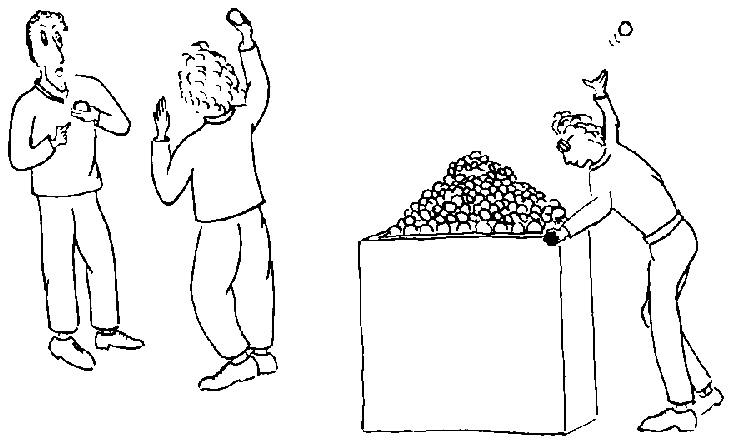}
\end{center}
\begin{center}
{\it He would simply pick and choose them.}
\end{center}
\end{figure}

\section{Hiawatha and the Experiments}

\begin{verse}
Hiawatha's revelation\\
Took a while to find its footing\\
For, as happens in such cases,\\
Many awful misconceptions\\
Were embedded in his thinking\\
Where they had been put on purpose\\
And could only be uncovered,\\
If at all, through painful hours\\
Scrutinizing tiny details,\\
Contemplating reams of data,\\
Finding out who's stuff was careful,\\
Tracking down suspicious rumors,\\
Reading through a mass of papers,\\
Slowly tossing out the bad ones,\\
Racking up the airline mileage\\
Going to humongous meetings,\\
Thereby building up a fact base\\
Cleansed of all manipulations.\\
Over time, as things got clearer,\\
Hiawatha grew unhappy\\
Trying to reconcile his viewpoint\\
With the facts that he had winnowed,\\
Always finding that he couldn't.\\
\end{verse}

\begin{verse}
Hiawatha studied transport\\
Both electrical and thermal\\
That, one argued, showed the absence\\
Of the Landau fermi surface\\
Symptomatic of a metal\\
Thereby proving one was dealing\\
With a strange new state of matter.\\
But he found in every instance\\
That a sample made its coldest,\\
So one knew what one was doing,\\
Either showed disorder problems\\
Generated by the chemists\\
Or agreed with classic theory.\\
Thus, like all those dot-com profits\\
That they claimed would make you wealthy,\\
But, in fact, were nonexistent,\\
Arguments for novel physics\\
Built upon the facts of transport\\
Did not hold up on inspection.\\
\end{verse}

\begin{verse}
Hiawatha studied optics\\
By and large his favorite spectrum\\
For he knew that light reflection\\
Measured dielectric functions\\
In a way that used no theory,\\
And it showed how loose electrons\\
Moved about and caused the bonding.\\
But, alas, the data varied\\
From one sample to another\\
Even after years of efforts\\
To ensure that they were stable!\\
This left lack of clear consensus\\
Even over things that mattered.\\
Understanding why this happened\\
Was not really rocket science,\\
For the Kramers-Kr\"{o}nig process\\
Amplified the defect signals\\
That were there in great abundance,\\
Even though they all denied it,\\
And depended on the process\\
By which one prepared the sample,\\
Something different for each grower\\
And a closely-guarded secret.\\
Also, things would change with doping,\\
Something very hard to measure\\
And which often wasn't constant\\
As one moved across the sample\\
Due to troubles in the furnace\\
Which they claimed they'd licked but hadn't.\\
Thus the stories of new physics\\
Built upon results of optics,\\
Like the troubled U. S. census\\
Or the the streets of downtown Boston\\
After weeks of too much snowing,\\
Were polluted by disorder,\\
And, moreover, were deceptive\\
In that aspects of the spectra\\
That were reasonably stable\\
Like the strange non-Drude lineshapes\\
Happened at such tiny wavelengths\\
One could plausibly ascribe them\\
To a nearby phase transition\\
Rather than the state in question.\\
Thus the stories were fantastic,\\
And, like those that Richard Nixon\\
Told while he was in the White House,\\
Or that pop star Michael Jackson\\
Claimed occurred in Los Olivos\\
For the pleasure of the children,\\
In the end would not hold water.\\
\end{verse}

\begin{figure}[t]
\begin{center}
\includegraphics{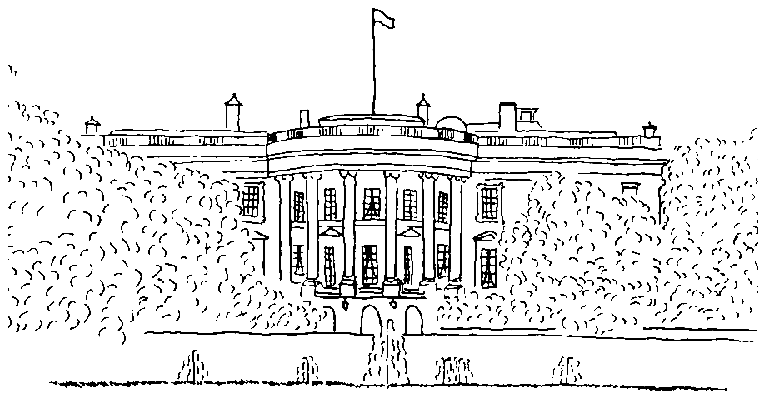}
\end{center}
\begin{center}
{\it Thus the stories were fantastic.}
\end{center}
\end{figure}

\begin{verse}
Hiawatha studied neutrons\\
Which he found he liked immensely\\
Since they flowed from a reactor\\
With big purple signs upon it\\
Warning you of radiation\\
That would kill you if allowed to,\\
Since the neutrons went right through you\\
But would sometimes choose to stop there\\
And decay like little time bombs,\\
Thus inducing stomach cancer.\\
But they went through cuprates also,\\
And that made them very useful,\\
Since a few of them would scatter,\\
And detecting those that did so\\
Gave you lots of information\\
From down deep inside the sample,\\
Such as how the atoms ordered,\\
How they moved when something hit them\\
And if they were little magnets.\\
But the bad news was the signal\\
Was quite small and hard to measure,\\
So one needed a detector\\
Bigger than a Dempsey Dumpster\\
And a truly mammoth sample,\\
Leading to big compromises\\
In the sample growing process\\
They preferred deemphasizing\\
But one knew was wreaking havoc\\
On the meaning of the data.\\
They would also never tell you\\
What the measurement itself was,\\
Since the neutron kinematics\\
Made it sensitive to factors\\
Like the speed spread of the neutrons\\
And the tip of the detector\\
And the path on which one moved things\\
To survey deflection angles\\
That were messy and annoying,\\
So they'd first massage the data\\
Using big computer programs\\
To remove these nasty factors\\
And report the program output,\\
Representing you should trust it\\
Just because they were the experts.\\
But, of course, there were those upgrades\\
And the quiet little tweaking\\
That one always did at run time\\
That one never heard reported.\\
Once he caught these key omissions\\
Hiawatha got suspicious,\\
And quite quickly found the practice\\
Of reporting neutron spectra\\
In some secret custom units\\
Given names like ``counts'' to fool you,\\
Like those helpful content labels\\
Found on packs of sandwich slices\\
Listing salt and beef by-products,\\
Thus preventing one from telling\\
There was very poor agreement.\\
All this made a clearer picture\\
But it also meant the data\\
Like the air-brushed prints in {\it Playboy}\\
Were, in fact, manipulated,\\
And that many strange behaviors\\
Like the famous funny phonon\\
Dogma said was nonexistent\\
Got removed as standard practice\\
On the grounds they should not be there.\\
Thus his plan to use those spectra\\
To pin down the magnetism\\
Present sometimes in the cuprates\\
On account of all the errors\\
Ended up a dismal failure.\\
\end{verse}

\begin{figure}[t!]
\begin{center}
\includegraphics{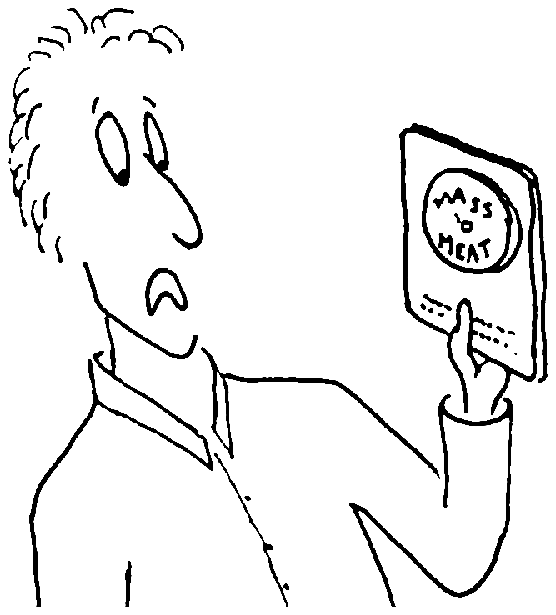}
\end{center}
\begin{center}
{\it Ever helpful content labels.}
\end{center}
\end{figure}

\begin{verse}
Hiawatha studied currents\\
Made when cold electrons tunnel\\
Right across an insulator\\
Where they should have been forbidden,\\
Something very close to magic\\
Rather like the twinkly transport\\
People undergo on Star Trek,\\
And it's also quite revealing\\
Of important quantum pairing\\
That goes on inside the cuprate.\\
In the old days one would simply\\
Oxidize a thin-film sample,\\
Coat the oxide with another,\\
Solder on two tiny contacts,\\
Dunk the whole thing into vapors\\
Made so cold that they were liquid,\\
Then just measure plain resistance\\
Of the two protruding wires,\\
Which would vary with the voltage\\
Thus producing useful data.\\
Hiawtha read these papers\\
With a mounting sense of horror,\\
For the wild disagreement\\
Even in the basic features\\
From one sample to another\\
Was so large it left one breathless.\\
And, of course, the accusations\\
That the other guys were morons\\
Who just could not make good junctions\\
Rose to unmatched heights of grandeur\\
Even though the real villain,\\
Obvious from spectral sharpness,\\
Was the sample variation.\\
Hiawatha's indignation\\
Escalated when he found that\\
Over time this fact got buried\\
Since each group soon found a method\\
Of preparing stable samples\\
Different from that used by others\\
And producing different spectra\\
That they marketed as products,\\
Thus evading any need to\\
Answer penetrating questions.\\
An important fact, however,\\
That emerged from all these studies,\\
Was that steady lossless currents\\
Could indeed be made to flow from\\
Films of lead into the cuprates\\
If one made a pitted surface,\\
Proving that the state of matter\\
Operating in the cuprates\\
Was not new and was not different.\\
\end{verse}

\begin{verse}
Hiawatha studied spectra\\
Made when light shined on a sample\\
Causes it to lose electrons\\
Which fly out in all directions\\
And one can detect by counting,\\
Thus obtaining information\\
Of their status in the sample\\
Just before the light removed them.\\
Hiawatha saw at once that\\
Peaks for plain undressed electrons\\
That were not supposed to be there\\
In this great new state of matter\\
Always were and had a sharpness\\
At the resolution limit\\
Of the latest new detector\\
For the special ones at threshold,\\
Where one knew what one was doing.\\
In addition they were beaming\\
In a lovely fourfold pattern\\
With the symmetry of d-wave,\\
Something that had been suggested\\
They might do if they were simple,\\
Just like those in other metals.\\
Thus the arguments for strangeness\\
Based on counting these electrons\\
Lost their force as things got better,\\
And in time were proved a failure.\\
\end{verse}

\begin{verse}
Hiawatha studied muons,\\
Which he thought were even neater\\
Than the more prosaic neutrons,\\
Since they came from atom smashers\\
That could also quickly kill you\\
If you chose to be so careless,\\
But they'd stop inside much better\\
And once there, decay to gammas\\
That were easily detected\\
Since they'd even go through concrete,\\
And, moreover, they'd be beaming\\
In the muon's spin direction\\
Just before it went to heaven.\\
Thus, implanted in a cuprate\\
They'd arrest at some location\\
Known to no one but their Maker\\
And precess like little searchlights,\\
If there was some magnetism,\\
Thus allowing you to see it\\
Way deep down inside the sample.\\
Thus with knowledge of their trapping\\
And a batch of big detectors\\
One could then back out the distance\\
Of magnetic penetration.\\
Hiawatha found this distance\\
Shortened with increasing doping\\
Just as theory said should happen,\\
If one forced the hubbard model\\
Not to be a quantum magnet\\
By just saying that it wasn't,\\
Which might well have been important\\
Had it not been for the problem\\
That this depth would not continue\\
To decline with increased doping\\
But instead would turn and lengthen.\\
This effect was quite perplexing,\\
Since no theory of the cuprates\\
Even twisted hubbard models,\\
Could account for such behavior,\\
For it violated sum rules,\\
Hence one just did not discuss it.\\
But the meaning was transparent\\
If one faced the facts with courage,\\
For the samples were degrading\\
In extremes of overdoping\\
In some ways that weren't predicted\\
And, moreover, weren't detected\\
By techniques except for this one.\\
This, in turn, implied these problems\\
Might occur at other dopings\\
And likewise escape detection\\
Or, what's worse, be used to argue\\
That new physics was occurring\\
When, in fact, it was just garbage.\\
Thus the trail blazed by muons\\
Led out in the woods to nowhere.\\
\end{verse}

\begin{figure}[t!]
\begin{center}
\includegraphics{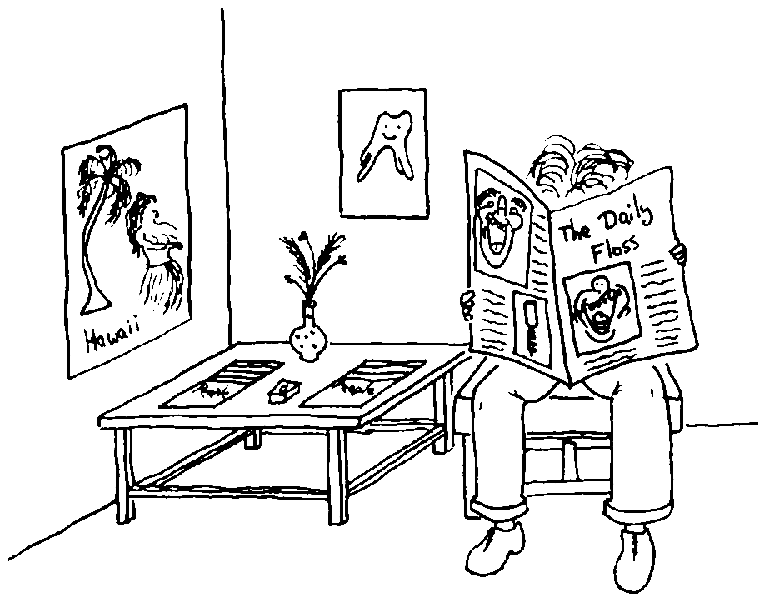}
\end{center}
\begin{center}
{\it Dental pamphlets make you tired.}
\end{center}
\end{figure}

\begin{verse}
Hiawatha studied spin flips\\
That the nuclei of atoms\\
Undergo in great big magnets\\
Near a radio transmitter\\
Causing them to be antennas,\\
Which absorb with complex lineshapes\\
One can read if one's a genius\\
But not, sadly, if one isn't,\\
Since they, by and large, consist of\\
Just a simple blobby bell curve\\
With a width and displaced center,\\
To which one must give some meaning---\\
Not a simple undertaking.\\
Thus the all-important Knight shift\\
And spin-lattice relaxation,\\
Noms de plume for width and center,\\
Vastly different for the copper\\
And the oxygen of cuprates,\\
Were the source of endless theories,\\
Often very thought-provoking,\\
Stunning in sophistication,\\
But, like all those glossy pamphlets\\
Found in waiting rooms of dentists\\
Urging you to practice flossing,\\
Soon began to make you tired,\\
Since the data mainly showed you\\
That the stuff was not a metal\\
In the sense of gold or iron\\
Which, in fact, one knew already\\
And was not a revelation.\\
\end{verse}

\begin{verse}
Hiawatha studied structure\\
Of the surfaces of cuprates\\
Freshly cleaved inside a vacuum\\
So that air would not get on them\\
And then probed with tiny needles\\
One could move with great precision,\\
By adjusting some piezos\\
On which everything was standing.\\
What he found was quite disturbing,\\
For while atoms at the surface\\
All had unperturbed positions,\\
Showing that the cleave succeeded,\\
There were also complex patterns\\
On the scale of twenty atoms\\
That appeared to be diffraction.\\
This behavior might have come from\\
Atoms underneath the surface\\
That were missing or defective\\
Or some novel magnetism\\
Of a kind unknown to science,\\
But the thing that so upset him\\
Was that quantum interference\\
Of the kind that he was seeing\\
Could not happen if the lifetimes\\
Were as short as he had thought them,\\
And which had been used to argue\\
For a brand new state of matter.\\
Thus he soberly concluded\\
That this matter wasn't different\\
And the whole confounded story\\
Was a misinterpretation\\
Of a plain materials problem.\\
\end{verse}

\begin{verse}
Thus the Mighty Hiawatha\\
Through the patient application\\
Of the practices of science\\
Tested over generations\\
Slowly sloughed off misconceptions\\
And, in face of mounting failure,\\ 
Sadly came to the conclusion\\
He'd been taken to the cleaners.\\
\end{verse}

\section{Hiawatha Befriends the Robots}

\begin{verse}
Given all the clever swindles\\
Lurking there to take our money,\\
That, of course, are part of living,\\
Like a virus for pneumonia\\
Or a hungry venus fly trap,\\
We must all be very thankful\\
That the celebrated Law of Murphy\\
Strikes at random without warning\\
Causing even brilliant concepts,\\
That appear completely foolproof\\
Like distributing tobacco\\
Or the business plan of Enron,\\
To sometimes become derailed\\
Due to something unexpected\\
One was sure could never happen,\\
Like a lawsuit from consumers,\\
That requires intervention\\
Of the most creative nature\\
To prevent strategic meltdown.\\
\end{verse}

\begin{figure}[t!]
\begin{center}
\includegraphics{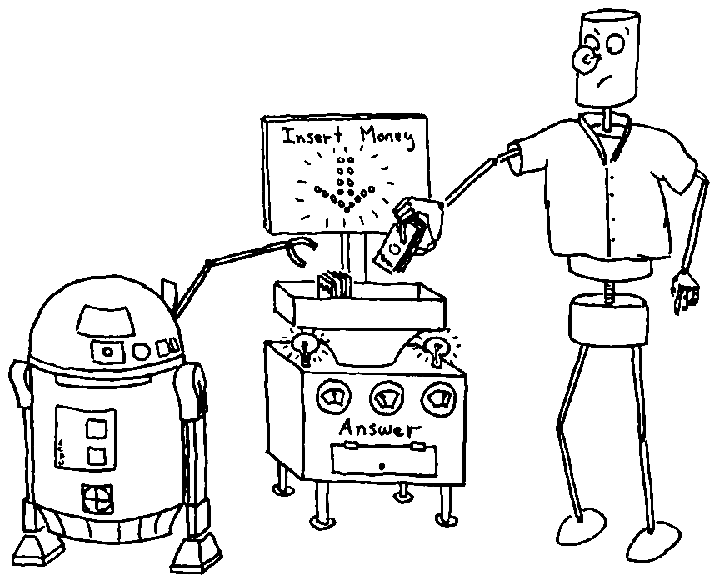}
\end{center}
\begin{center}
{\it Sure enough, that's just what happened.}
\end{center}
\end{figure}

\begin{verse}
As it turns out, the idea\\
That the conflict in the models\\
One was using for the cuprates\\
Due to nearby phase transitions\\
Would both hamper their solution\\
And engender rampant fibbing,\\
Thus enshrining mass confusion\\
One would then call proof of meaning\\
With no need to fear exposure\\
Had the unexpected weakness\\
That someone might {\it solve} the model\\
Using tons and tons of money\\
And some capable computers\\
To a crude degree sufficient\\
To unmask the real problem\\
Thus revealing the deception.\\
\end{verse}

\begin{verse}
Sure enough, that's just what happened.\\
When the cuprates were discovered\\
And the whole endeavor started\\
One had not the slightest worry\\
That these guys would ever solve it,\\
Since the accuracy needed\\
Was not clear in the beginning,\\
So they uniformly low-balled\\
With the too-familiar outcome\\
That results were inconsistent.\\
So they quarrelled over method\\
And who had convergence problems\\
And whose code was most clairvoyant\\
Even though a child could see that\\
They were different apparati,\\
So the test that they were working\\
Was agreement with each other.\\
But, unlike the other issues\\
That had come and gone before it,\\
Cuprates lingered on as timely\\
Long enough to cause a shake-out,\\
For the money kept increasing\\
Even as machines got cheaper\\
And their power kept on growing---\\
Due, of course, to needs of gaming,\\
Rather than the ones of Lanczos\\
Or the quantum monte carlo\\
That one used for basic physics.\\
So the robots kept on plugging\\
As their owners upped the ante\\
Very slowly, as did Wagner\\
When composing {\it Ring} and {\it Tristan}\\
And their stuff began converging!\\
There, of course, was no agreement\\
Over matters of the phases\\
Such as whether it conducted\\
When one cooled it down to zero,\\
Since a crystal of electrons\\
Was one state in competition.\\
But at length scales one could access\\
There was clearly dissipation\\
Of a most peculiar nature\\
In the dielectric function\\
And the quantum magnetism,\\
Just exactly as predicted\\
By an ancient bunch of papers\\
Over quantum phase transitions,\\
Which these guys had never studied\\
Since it was too esoteric\\
And had not been seen in nature\\
And was hated by their funders.\\
But the thing that really clinched it\\
Was the endless disagreement,\\
That got worse as things proceeded\\
And was very clearly cronic,\\
Over type and shape of edges\\
That would best produce convergence,\\
Since one found that subtle changes\\
In the way one built the model\\
Would turn on and off the striping\\
And therefore the insulation,\\
So that whether it was present\\
In the limit of large sample\\
Simply could not be determined\\
With the codes that they had written.\\
\end{verse}

\begin{figure}
\begin{center}
\includegraphics{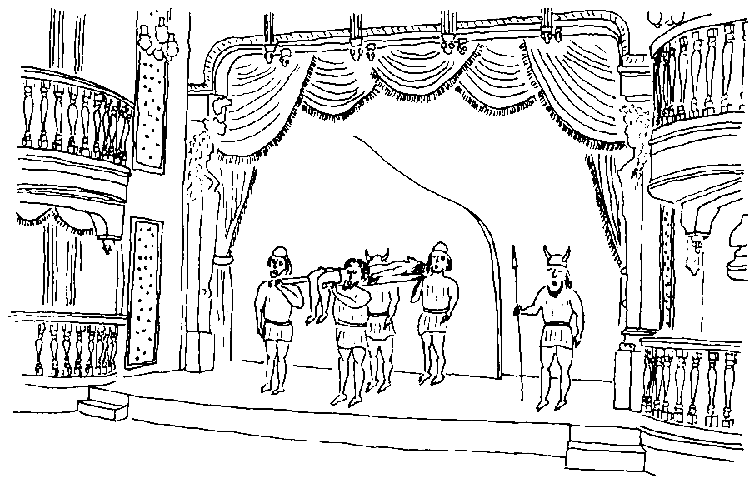}
\end{center}
\begin{center}
{\it Soon their stuff began converging. }
\end{center}
\end{figure}

\begin{verse}
This, of course, was a disaster\\
For the plan to keep things murky\\
And required drastic action\\
To somehow repair the damage\\
All this progress had created,\\
And prevent these guys from seeing\\
What was right beneath their noses.\\
\end{verse}

\begin{verse}
And one was not disappointed.\\
Once again a flash of brilliance\\
Like a great big city-buster\\
Brighter than the sun at midday,\\
Blazed across the dome of heaven\\
Toward its final destination\\
In the Guinness Book of Records.\\
They declared the problem {\it over}!\\
The computer guys had solved it!\\
For their codes had proved the cuprates\\
Were indeed the Hubbard model,\\
And that's why the stuff conducted.\\
Thus there was no urgent reason\\
To pursue the matter further!\\
One could zero out their budgets\\
With no loss to human knowledge\\
And, in fact, perhaps improve it\\
Since this money was incentive\\
To continue calculations\\
That were clearly unimportant\\
And report them in the journals\\
Thus just adding to the clutter.\\
\end{verse}

\begin{verse}
Hiawatha, now much wiser\\
Through his labors as a scholar\\
And, quite frankly, some maturing\\
Watched these things unfold before him,\\
As he had on past occasions,\\
But this time with eyes wide open\\
And was filled with understanding.\\
It was not a happy moment,\\
For it meant that his own judgement\\
As to what was good and worthy\\
Had been faulty from the outset,\\
Something for which he must answer.\\
But instead of indignation\\
And a passion to get even\\
That he might have felt when younger\\
Hiawatha, deep in thinking,\\
Found himself consumed with sadness.\\
He was not the only victim,\\
For the guys who manned those robots,\\
And were heroes of the cuprates---\\
For through focussed dedication\\
They had stumbled on the answer\\
That the models were unstable\\
And did {\it not} describe the cuprates,\\
Since a modest perturbation\\
Would profoundly change their nature---\\
Were about to have their triumph\\
Snatched from them by clever scoundrels\\
Who, pretending to befriend them,\\
Would then redefine their output\\
To mean something that it didn't,\\
Thus protecting their investment,\\
But, of course, destroying others.\\
\end{verse}

\section{Hiawatha's Lamentation}

\begin{verse}
Hiawatha's knowing sadness,\\
Like the darkening at twilight\\
Or a gathering storm in winter,\\
Slowly gained in strength and deepened\\
As he spent time in reflection,\\
Working through the implications\\
Of the things that he had witnessed\\
For the cause of noble science\\
That thus far had so beguiled him.\\
It would simply not be manly\\
To pretend he wasn't guilty\\
Of ignoring frequent warnings\\
That the needed path to nature\\
Was obscured or nonexistent.\\
It was clear that he'd been foolish\\
To have bought this awful fiction\\
And that blame must fall quite squarely\\
On himself and not on others.\\
But this candid {\it mea culpa},\\
Made in silence where it mattered,\\
While it comforted his conscience,\\
Did not quite assuage the wounding,\\
For it begged the nagging question\\
Of how they could have succeeded\\
In hoodwinking all the people\\
For so long without some doubting.\\
It was simply not an option\\
To presume these guys were stupid,\\
Since the instruments they dealt with,\\
Often built by hand from nothing,\\
Needed great sophistication\\
To deploy and mine for data.\\
There was clearly something larger\\
And extremely fundamental\\
Working in the group dynamic\\
That involved access to funding\\
And the policy of journals\\
And the need to service markets\\
And the mythos of the subject\\
One must use to make a living\\
That these crooks had first deciphered,\\
Then reduced with understanding,\\
Then usurped to do their bidding.\\
\end{verse}

\begin{figure}[t!]
\begin{center}
\includegraphics{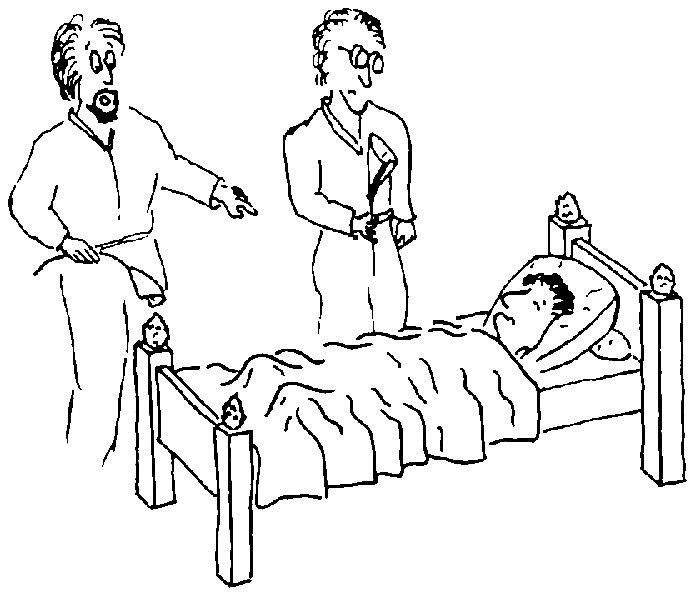}
\end{center}
\begin{center}
{\it Then the body gets diseases.}
\end{center}
\end{figure}

\begin{verse}
Hiawatha, turning inward,\\
Thought for weeks about this problem\\
During which he was obnoxious\\
Due to his preoccupation.\\
But at last he got an answer\\
That made sense and was quite simple,\\
Thus withstanding Occam's razor,\\
So he thought that he believed it.\\
When he'd set out on his mission\\
He had understood the challenge\\
Of the mastery of nature\\
But not basic economics\\
And the fact that art and science\\
Both require sacrifices\\
Of a clear financial nature\\
That one sometimes just can't handle\\
Nor, in fairness, should one do so\\
Since a good guy pays the mortgage\\ 
And supports the kids in college\\
And the other things a body\\
Has to do to keep the lights on.\\
But, in fact, the compromises\\ 
That one makes as part of living\\
Such as saying what one has to\\
For maintaining healthy cash flow\\
Often toss big monkey wrenches\\
In the fine machine of science\\
And can stop it altogether\\
In conflicted situations.\\
Then the body, badly weakened,\\
Barely able to keep breathing,\\
Gets exploited by diseases,\\
Such as villains lacking scruples\\
Who descend on it like termites\\
To a house that's been neglected,\\
Wreaking terrible destruction\\
On the lives of those affected.\\
\end{verse}

\begin{verse}
The conclusion of this story\\
Is well known from all the textbooks.\\
Hiawatha never wavered\\
In his deep respect for physics,\\
But he came by this adventure\\
To the deeper understanding\\
That to get things done that mattered\\
Often was a social question,\\
Not just logical abstraction,\\
And, as well, a part of nature,\\
Just the thing he thought he'd hated\\ 
And had thrilled at desecrating\\
As a tender freshman student\\
In the little private college\\
By the shores of Gitchee-Gumee.\\
It was true that all the creatures\\
Living in those swamps and woodlands\\
Generated lots of pooping,\\
But then so did real people,\\
And the people poop was stronger,\\
So that one could not ignore it.\\
But one really would not want to,\\
For the lesson of the cuprates\\
Was that lack of understanding\\
Of these basic group dynamics,\\
Was a recipe for failure\\
Since they were the central issue\\
For most things that were essential.\\
\end{verse}

\begin{figure}[t!]
\begin{center}
\includegraphics{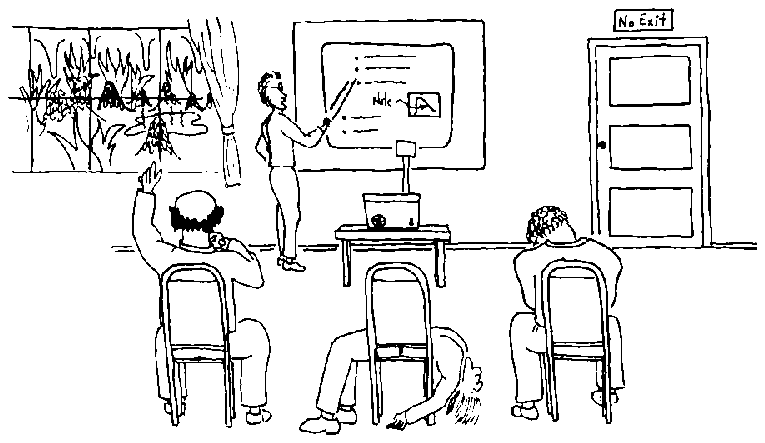}
\end{center}
\begin{center}
{\it They are forced to sit through lectures.}
\end{center}
\end{figure}

\begin{verse}
Thus the mighty Hiawatha\\
Turned his mind to other problems\\
Such as how to use resources\\
That were his by luck and birthright\\
Through the power of his father\\
Which he'd been inclined to squander,\\
But now realized he shouldn't.\\
Thus he studied like a madman\\
To acquire the skills of statecraft,\\
Such as how to plan a project,\\
How to give effective orders,\\
How to make sure they were followed,\\
How to get things done with meetings,\\
And to leave the money grubbing\\
Up to folks his father hired\\
Such as all those gifted spin docs\\
Who created key revisions\\
Necessary for his image\\
To be something people honored.\\
Thus the pain of too much sliding\\
On the ice in dead of winter\\
In an inexpensive loincloth\\
And his other misadventures\\
Got removed, as did the cuprates,\\
From his long official story.\\
But the memory persisted\\
And it helped to make him wiser\\
For, of course, as he got older\\
He had many bad encounters\\
Not so different from the cuprates.\\
\end{verse}

\begin{verse}
But whenever he was troubled\\
With a problem that would vex him\\
He would cheer himself by thinking\\
Of the special room in Hades\\
Into which these happy people\\
On account of their transgressions\\
Would be ushered when they bagged it\\
And be stuck in there forever,\\
Forced to listen to each other\\
Giving lectures on the cuprates.\\
It would always leave him smiling.\\
\end{verse}

\end{document}